\documentclass[11pt]{article}
\usepackage[margin=1in]{geometry}

\usepackage{authblk}
\usepackage{amsmath,amssymb,amsthm}
\usepackage{mathtools}
\usepackage{url}

\newcommand{\rel}{\mathbb{R}}

\newcommand{\nat}{\mathbb{N}}

\newcommand{\st}{\colon}

\newtheorem{lem}{Lemma}

\title{On Deciding the Constraint Loop Problem in Dimension Two\footnote{This is a report of the results produced by the first author while she was a research intern at the MPI-SWS, supervised by the second author.}}
\author[2]{Daria Goliasova}
\author[1]{Toghrul Karimov}
\affil[1]{Max Planck Institute for Software Systems, Saarbr{\"u}cken, Germany}
\affil[2]{Virginia Tech, Blacksburg, Virginia, USA}
\date{\today}

\begin{document}
\maketitle

\begin{abstract}
    Recently, Guilmant, Lefaucheux, Ouaknine, and Worrell \cite{guilmant_et_al:LIPIcs.ICALP.2024.140} proved decidability of the \emph{termination problem for linear constraint loops} over two real-valued variables, thus settling a well-known open problem. 
    Their algorithm is based on a theoretical result that such a loop does not terminate if and only if it has a witness of an infinite run of a specific shape.
    By analysing the structure of the witness carefully, we show that termination such loops can be decided in polynomial time, improving on the polynomial space bound argued by Guilmant et al.
    We also present an efficient implementation of the polynomial-time algorithm based on the satisfiability modulo theories (SMT) solver \texttt{Z3}.
\end{abstract}

\section{Introduction}

The problem of program verification is essential in modern software engineering. 
Ensuring that a program executes safely without getting stuck in infinite execution requires proving \emph{loop termination}.
In this work, we focus on termination for \emph{non-deterministic linear constraint loops}, which has been extensively studied: see \cite{ben2026termination} for a recent survey. 
In this setting, both the loop guard and the loop update are specified as conjunctions of linear inequalities over program variables:
\[
\textrm{while} \:\: B \cdot x \ge b \:\: \textrm{do} \:\: A \cdot (x,  x') \ge a
\]
where $x, x'$ are respectively the column vectors of program variables before and after the loop iteration. $A, B$ are matrices with rational entries, $a,b$ are vectors of rational numbers.
By folding the guard condition into the update we can write every such loop as
\[
\textrm{while} \:\: \textrm{True} \:\: \textrm{do} \:\: A \cdot (x, x') \ge b.
\]
The termination question (over $\rel$) is then equivalent to existence of an infinite sequence $(x_n)_{n \in \nat}$ of vectors of real numbers such that $A \cdot (x_n, x_{n+1}) \ge b$ for all $n \in \nat$.

Given $(A, b)$ as above, we define
\[
K = \{(x, x') \colon \textrm{$x$ to $x'$ is a valid transition of the loop. i.e.\ $A \cdot (x,x') \ge b$}\}.
\]
The set $K$ contains all information about the loop.
When $x, x' \in \rel^2$, i.e.\ the program has two variables and $K \subseteq \mathbb{R}^4$, recently \cite{guilmant_et_al:LIPIcs.ICALP.2024.140} gave an algorithm for deciding termination.
Their argument is that the loop does not terminate if and only if there exists a witness as below; from the witness it is then easy to extract an explicit infinite trajectory.
A witness consists of a linear map $M \colon \mathbb{R}^2 \to \mathbb{R}^2$, a closed convex cone $C \subseteq \mathbb{R}^2$, and vectors $v, w \in \mathbb{R}^2$ satisfying four conditions:
\begin{itemize}
    \item $M \cdot C \subseteq C$,
    \item $\forall z \in C \colon (z, Mz) \in \operatorname{rec}(K)$,
    \item $(v,w) \in K$,
    \item $w - v \in C$.
\end{itemize}
Here
\[
\operatorname{rec}(K) = \{z \colon w + cz \in K \textrm{ for all } c \ge 0 \textrm{ and } w \in K \}
\]
is the \emph{recession cone} of $K$.
The witnessing infinite trajectory is then \cite[Prop.~27]{guilmant_et_al:LIPIcs.ICALP.2024.140}
\[
v, \quad w, \quad v + M(w-v), \quad v + M(w-v) + M^2(w-v), \quad \dots
\]
So we have the partial sums of the geometric sequence $(M^n(w-v))_{n \in \nat}$ translated by $v$; this form of a witnessing infinite trajectory was already known in \cite{leike2018geometric}.

The original goal of this project was to implement an automated verification tool in Python using the \textbf{Z3 SMT solver} to search for such non-termination witnesses. 
However, we quickly realised that the naive translation of the requirements into first-order logic requires quantifier alternations (albit with a fixed number of variables) that Z3 cannot handle. 
In fact, Guilmant et al.\ note the complexity of checking existence of a witness as PSPACE, which comes from the aforementioned naive translation. 
We therefore first show how to decide existence of a witness in polynomial time, and then implement that in Z3.
The implemented tool is available at \url{loop-termination.mpi-sws.org}.

\section{Checking existence of a witness in polynomial time}

We perform a case analysis based on $C$.
The polynomial running time follows from the result of \cite[Thm.~1.1]{renegar1992computational} that for any fixed $k$, deciding truth of $\exists x_1, \ldots, x_k \colon \varphi(x_1,\ldots,x_k)$, where $\varphi$ is Boolean combination of polynomial (with integer coefficients) inequalities in $x_1,\ldots,x_k$, can be done in polynomial time. 
Here the exact polynomial bound on the number of arithmetic operations does depend on $k$.

Firstly, let us consider witnesses $(M, C, v, w)$ where $C = \{ c_1 s + c_2 t \colon c_1, c_2 \ge 0\}$ for two boundary vectors $s$ and $t$ that may be zero: in this case $C$ is either a two-dimensional pointed cone, a ray, a line, or a singleton.

\subsection{Encoding the witness conditions}
The program formulates four sets of constraints corresponding to the four witness conditions of~\cite{guilmant_et_al:LIPIcs.ICALP.2024.140}.
To get polynomial time, what needs to be done is to not go for the direct transcription of the property into first-order logic (which may involve quantifier alternations), and instead use a mathematical argument.

\begin{enumerate}
    \item $M \cdot C \subseteq C$. The generator vectors $s$ and $t$ transformed by matrix $M$ must remain inside $C$:
\[
\operatorname{belongs\_to\_C}(Ms, s, t) \quad \land \quad \operatorname{belongs\_to\_C}(Mt, s, t)
\]
To see that this is valid, suppose $M s = a_1 s + a_2 t$ and $M t = b_1 s + b_2 t$ for $a_i, b_i \ge 0$.
Take a vector $z = c_1 s + c_2 t \in C$ where $c_1,c_2 \ge 0$.
Then
\[
M z = (c_1a_1 + c_2b_1)s + (c_1a_2 + c_2b_2)t.
\]
Since $c_1a_1 + c_2b_1, c_1a_2 + c_2b_2 \ge 0$, we have that $M z \in C$.
    \item $\forall z \in C \colon  (z, Mz) \in \operatorname{rec}(K)$. 
    To eliminate the universal quantifier, we only check that the generators of $C$ satisfy the recession cone condition:
    \[
    (s, Ms) \in \operatorname{rec}(K) \quad \land \quad (t, Mt) \in \operatorname{rec}(K).
    \]
    This is justified as follows.
    Assume the condition for $s$ and $t$.
    Take $z = c_1s + c_2t \in C$ with $c_1, c_2 \ge 0$.
    Then, applying Lem.~\ref{rec},
    \[
    A \cdot (z, Mz) = A \cdot ((c_1s, c_1Ms) + (c_2t, c_2Mt)) = c_1A \cdot (s, Ms) + c_2A \cdot (t, Mt) \ge 0.
    \]
    From Lem.~\ref{rec} it follows that $(z, Mz) \in \operatorname{rec}(K)$.
    \item $(v,w) \in K$. This does not require any quantifiers, and can be written in the straightforward way as a Z3 formula. 
    This condition enforces that state $v$ satisfies the loop guard and transition $(v, w)$ obeys loop constraints.
    \item $w-v \in C$. This enforces that the step vector $w-v$ falls within the cone $C$, and can be encoded as
    \[
    \operatorname{belongs\_to\_C}(w - v, s, t).
    \]
\end{enumerate}

The global formulas to be checked are then of the form $\exists s, t, v, w, M \st \varphi$ where $\varphi$ is a quantifier-free formula containing the usual logical operations as well as $\operatorname{rec}(K)$ and $\operatorname{belongs\_to\_C}$.
Next we will show how to implement these using quantifier-free Z3 formulas (that use non-linear arithmetic).

\subsection{Testing membership in a given cone and the recession cone}

First we show how to check $z \in \operatorname{rec}(K)$ using no quantifiers and just linear operations.
\begin{lem}
    \label{rec}
    Assuming $K \ne \varnothing$, we have that $z \in \operatorname{rec}(K)$ if and only if $A z \ge 0$ where $\ge$ works entrywise on vectors.
\end{lem}
\begin{proof}
    Suppose $A z \ge 0$.
    Take $w \in K$ and $c \ge 0$.
    Then
    \[
    A (w + cz) = A w + c A z \ge Aw \ge b
    \]
    and hence $z \in \operatorname{rec}(K)$.

    Now suppose $z \in \operatorname{rec}(K)$.
    Let $a_i$ be a row of $A$, viewed as a column vector, and $b_i$ be the corresponding entry of $b$.
    We have to show that $a_i \cdot z \ge 0$.
    Take any $w \in K$.
    By assumption for any $c \ge 0$ we have that $a_i \cdot (w + cz) = a_i \cdot w + c(a_i \cdot z) \ge b_i$.
    This cannot hold if $a_i \cdot z < 0$.
\end{proof}

Let us move on to defining $\operatorname{belongs\_to\_C}$.
A vector $x = (x_1, x_2)$ belongs to the cone $C$ generated by $s,t$ if it can be represented as
\[
x = c_1 s + c_2 t, \quad \operatorname{where } c_1 \ge 0, \; c_2 \ge 0.
\]
Solving this system for $c_1$ and $c_2$ using Cramer's rule yields
\[
c_1 = \frac{x_1 t_2 - x_2 t_1}{s_1 t_2 - s_2 t_1}, \qquad 
c_2 = \frac{s_1 x_2 - s_2 x_1}{s_1 t_2 - s_2 t_1}.
\]
To avoid division inside Z3 constraints, we clear the denominator $D = s_1 t_2 - s_2 t_1$. Since the order of the generators $(s, t)$ is arbitrary (recall that both $s$ and $t$ will be existentially quantified), we enforce $D \ge 0$ without loss of generality.

\begin{description}
    \item[Case $D > 0$.] In this case $x \in C$ is equivalent to
    \[
    x_1 t_2 - x_2 t_1 \ge 0 \quad \land \quad s_1 x_2 - s_2 x_1 \ge 0.
    \]
    \item[Case $D = 0$.] Vectors $s$ and $t$ are collinear. We check whether $x$ lies directly on the ray defined by $s$ and $t$.
\end{description}

\noindent Thus in Z3, $\operatorname{belongs\_to\_C}(x, s, t)$ is formulated as
\begin{align*}
    &D \ge 0 \\
    &\big( D > 0 \implies (x_1 t_2 - x_2 t_1 \ge 0 \land s_1 x_2 - s_2 x_1 \ge 0) \big) \quad\land\quad \\ 
    &\big( D = 0 \implies (\operatorname{on\_ray}(x, s) \lor \operatorname{on\_ray}(x, t)) \big) \quad\land\quad 
\end{align*}
where $\operatorname{on\_ray}(x, y)$ (meaning $x$ belongs to the ray generated by $y$) for $y = (y_1,y_2)$ is
\[
(y = 0 \Rightarrow x = 0) \quad\lor\quad (y \ne 0 \:\land\: x_1y_2 = x_2y_1 \:\land\: x_1y_1 \ge 0 \:\land\: x_2y_2 \ge 0)
\]
which belongs to the non-linear real arithmetic fragment.
Note that the formula is quantifier-free.

\subsection{Different type of witness check}

The witness includes two vectors that form a cone. 
However, the two vector cone might not be necessary for working witness. The cone could have simpler form. The program would have a shorter running time if additional information was known about \textbf{s} and \textbf{t}.

The program does a check to find a witness with $C$ in form of point, ray, line, cone. To ensure the result is a cone of the required form, the additional constraints are added to the solver, on top of witness constraints:

\begin{itemize}
    \item point:
        $s = t = 0$
        
    \item ray:
        $s = t, s \ne 0$
    
    \item line:
        $s = -t, s \ne 0$
    
    \item non-degenerate two-dimensional cone:
        $s \neq t, s \neq -t, s\ne 0, t \ne 0$
\end{itemize}

These checks ensure that the cone would have criteria for required form, but also would not be of simpler form than required.

\subsection{The case of a half-plane $C$}

A separate type of possible cone is a half-plane, which is generated by three rather than two vectors.
However, the three vectors generating such a cone can be taken from the restricted family $s \coloneqq (s_1, s_2) \ne 0, t \coloneqq -s, u \coloneqq (-s_2, s_1)$: note that the latter is orthogonal to $s$.
The algorithm had to be changed to properly check different parts of witness for the half-plane cone.

\begin{enumerate}
    \item $M \cdot C \subseteq C$. The generator vectors $s$, $t$ and $u$ transformed by matrix $M$ must remain inside $C$, therefore they have to be inside one of cones generated by $s$ and $u$ or $u$ and $t$:
    \begin{align*}
        &(\operatorname{belongs\_to\_C}(Ms, s, u) \quad \lor \quad \operatorname{belongs\_to\_C}(Ms, u, t)) \quad\land\quad \\
        &(\operatorname{belongs\_to\_C}(Mu, s, u) \quad \lor \quad \operatorname{belongs\_to\_C}(Mu, u, t)) \quad\land\quad \\
        &(\operatorname{belongs\_to\_C}(Mt, s, u) \quad \lor \quad \operatorname{belongs\_to\_C}(Mt, u, t))
    \end{align*}
    \item $\forall z \in C \colon  (z, Mz) \in \operatorname{rec}(K)$. 
    As argued above, it suffices to reason only about the three generators:
    \[
    (s, Ms) \in \operatorname{rec}(K) \quad \land \quad (u, Mu) \in \operatorname{rec}(K) \quad \land \quad (t, Mt) \in \operatorname{rec}(K)
    \]
    \item $(v,w) \in K$. Same as before.
    \item $w-v \in C$. This time we have
    \[
    \operatorname{belongs\_to\_C}(w - v, s, u) \quad \lor \quad \operatorname{belongs\_to\_C}(w -v, u, t)
    \]
    since the cone generated by $s, t, u$ is the union of the two cones generated by $s,u$ and $u, t$.
\end{enumerate}

\subsection{Prechecks}
To increase the program run speed, before running the witness check, program checks if the polytope formed by constraints is not empty and is not bounded. 
These are done using ordinary linear programming.
In case of the polytope being empty no further checks are needed, as the loop would terminate after one iteration for any initial value. 
When the polytope is bounded, the loop does not terminate if and only if it has a fixed point \cite[Prop.~26]{guilmant_et_al:LIPIcs.ICALP.2024.140}.

\section{Our implementation}

The program is implemented in Python using the SMT solver library \texttt{z3-solver}.
The exact linear programming on Python integers (which do not have a size restriction) as well as \texttt{Fraction}s is performed using \texttt{pycddlib}.
Table~\ref{tab:constraints_time} shows the running time of our program as a function of the number of constraints: the $\pm$ indicates one standard deviation of the distribution of running times.
In all experiments, $A$ had entries of absolute value 100, and the constraints were generated randomly but in a controlled way so that $K$ was unbounded and non-empty: otherwise the pre-checks would have stopped the execution.
For each number of constraints, 300 instances were constructed.
We see that our program is able to handle a very large number of constraints, even if it takes up to a minute.

The code can be found in \url{https://github.com/DariaGoliasova/loop-termination}

The online interface \url{loop-termination.mpi-sws.org} was implemented using Flask in Python. 
Based on the data from our experiment, it has a timeout limit of 10 seconds.

\begin{table}
    \centering
    \begin{tabular}{|r|r|}
        \hline
        \textbf{Number of constraints} & \textbf{Approximate time (s)} \\
        \hline
        100 & 7.066 ± 6.099 \\
        1000 & 9.060 ± 5.832 \\
        10000 & 43.570 ± 3.369 \\
        \hline
    \end{tabular}
    \caption{Approximate execution time by number of constraints}
    \label{tab:constraints_time}
\end{table}

\section{Open problems}

A natural question that arises from checking witnesses with simpler "cones" is whether loops sharing a specific solution type--a point, ray, line, half-plane, full-plane or no simpler solution--possess common underlying characteristics. Furthermore, is there a method to determine which type of solution exists for a given loop a priori, without having to construct it explicitly?

A related challenge concerns how the algorithm behaves in higher dimensions. Developing more efficient search or verification strategies for higher dimensions remains an important direction for future work: currently there is no theoretical result for three variables, and even checking membership in a cone generated by many extremal vectors (in 3D, unlike in 2D, a cone need not be generated by a fixed number of vectors) already causes implementation difficulties.

\bibliographystyle{plain}
\bibliography{refs}

\end{document}